\pdfoutput=1
\documentclass[fleqn,usenatbib]{mnras}

\usepackage{newtxtext}

\usepackage[T1]{fontenc}

\DeclareRobustCommand{\VAN}[3]{#2}
\let\VANthebibliography\thebibliography
\def\thebibliography{\DeclareRobustCommand{\VAN}[3]{##3}\VANthebibliography}

\usepackage{graphicx}	
\usepackage{amsmath}	
\usepackage{amssymb}	
\usepackage[utf8]{inputenc}
\usepackage[english]{babel}

\usepackage[dvipsnames]{xcolor}
\definecolor{mypink1}{RGB}{255, 127, 127}
\definecolor{mypurple}{RGB}{179, 0, 255}

\title[ ]{Wave Dark Matter and Ultra Diffuse Galaxies.}

\author[A. Pozo et al.]{
Alvaro Pozo,$^{1,2}$\thanks{E-mail: mn@ras.org.uk (KTS)}
Tom Broadhurst,$^{1,2,3}$
Ivan de Martino,$^{2,4,5}$
Hoang Nhan Luu,$^{6}$
\newauthor{
George~F.~ Smoot,$^{6,7,8,9}$
Jeremy Lim,$^{6}$
Mark Neyrinck$^{1,2,3}$
}
\\
$^{1}$Department of Physics, University of the Basque Country UPV/EHU, E-48080 Bilbao, Spain;\\ email:alvaro.pozolarrocha@bizkaia.eu; tom.j.broadhurst@gmail.com;\\
$^{2}$DIPC, Basque Country UPV/EHU, E-48080 San Sebastian, Spain\\
$^{3}$Ikerbasque, Basque Foundation for Science, E-48011 Bilbao, Spain\\
$^{4}$Dipartimento di Fisica, Universit\`a di Torino,  Via P. Giuria 1, I-10125 Torino, Italy; ivan.demartino@unito.it\\
$^{5}$Istituto Nazionale di Fisica Nucleare (INFN), Sezione di Torino, Via P. Giuria 1, I-10125 Torino, Italy\\
$^{6}$Institute for Advanced Study and Department of Physics, Hong Kong University of Science and Technology, Hong Kong\\
$^{7}$Institute for Advanced Study, The Hong Kong University of Science and Technology, Clear Water Bay, Kowloon, Hong Kong\\
$^{8}$AS TT \& WF Chao Foundation Professor, IAS, Hong Kong University of Science and Technology, Clear Water Bay, Kowloon, 999077 Hong Kong\\
$^{9}$Paris Centre for Cosmological Physics, APC, AstroParticule et Cosmologie, Universit\'{e} Paris Diderot,CNRS/IN2P3, CEA/lrfu, \\
Universit\'{e} Sorbonne Paris Cit\'{e}, 10, rue Alice Domon et Leonie Duquet, 75205 Paris CEDEX 13, France
}

\date{22 March 2021}

\pubyear{2020}

\begin{document}
\label{firstpage}
\pagerange{\pageref{firstpage}--\pageref{lastpage}}
\maketitle

\begin{abstract}
Dark matter as a Bose-Einstein condensate, such as the axionic scalar field particles of String Theory, can explain the coldness of dark matter on large scales. Pioneering simulations in this context predict a rich wave-like structure, with a ground state soliton core in every galaxy surrounded by a halo of excited states that interfere on the de Broglie scale. This de Broglie scale is largest for low mass galaxies as momentum is lower, providing a simple explanation for the wide cores of dwarf spheroidal galaxies. Here we extend these ``wave dark matter" ($\psi$DM) predictions to the newly discovered class of ``Ultra Diffuse Galaxies" (UDG) that resemble dwarf spheroidal galaxies but with more extended stellar profiles. Currently the best studied example, DF44, has a uniform velocity dispersion of $\simeq 33$km/s, extending to at least 3 kpc, that we show is reproduced by our $\psi$DM simulations with a soliton radius of $\simeq 0.5$ kpc. In the $\psi$DM context, we show the relatively flat dispersion profile of DF44 lies between massive galaxies with compact dense solitons, as may be present in the Milky Way on a scale of 100pc and lower mass galaxies where the velocity dispersion declines centrally within a wide, low density soliton, like Antlia II, of radius 3 kpc. 
\end{abstract}




\section{Introduction}

The origin of Dark Matter (DM) is understood to lie ``beyond" the standard model of particle physics, which describes only 16\% of the total cosmological mass density  \citep{Cyburt2016,Planck16}. It is also clear that DM is non-relativistic, to the earliest limits of observation given the Cosmic Microwave Background and galaxy power spectrum. Furthermore, DM behaves collisionlessly for pairs of clusters undergoing a collision, implying there is no significant self interaction other than gravity, in particular the iconic Bullet cluster \citep{Marikevitch2004,Clowe2006} and other massive examples \citep{MB2018}.

This collisionless ``Cold Dark Matter" (CDM) has long been synonymous with heavy particles of some new form, that must be unusually stable and interact only via gravity, and principally motivated by super-symmetric extensions of the Standard Model \citep{Ellis84}. 
However, enthusiasm for WIMP-based CDM is now fading with the continued absence of any such particle in stringent direct laboratory searches \citep{Aprile2018} and astrophysical DM annihilation searches \citep{Archambault2017} and because of long recognised tension in this context with the basic 
observed properties of low mass galaxies\citep{Moore1994,Bullock2017}.

The dwarf galaxy tension applies to the primordial black hole interpretation of DM, and has been revived by the puzzlingly strong LIGO events \citep{Bird2016}, as black holes effectively act as collisionless test particles, like WIMP's (except when coalescing). However, the LIGO inferred black hole mass (BBH) scale of $30M_\odot$ is not present in Galactic microlensing searches, nor in the light curves of individual lensed stars recently discovered through high columns of cluster DM, limiting any such PBH contribution to a small fraction of the DM \citep{Diego2018,Kelly2018,Oguri2018}.

Instead, a light boson solution for the universal DM is gaining credence on the strength of the first simulations to evolve the coupled Schr\"ordinger-Poisson equations, dubbed $\psi$DM, \citep{Schive2014} that reveal distinctive, testable predictions for the non-linear structure of this wave-like form of DM. The Uncertainty Principle means bosons cannot be confined to a scale smaller than the de Broglie one, so this ``pressure" naturally suppresses dwarf galaxy formation and  generates rich structure on the de Broglie scale, as revealed by the first simulations \citep{Schive2014,Schive2014b,Mocz2017,Veltmaat2018} in this context. The most distinctive $\psi$DM prediction is the formation of a prominent solitonic standing wave at the base of every virialised potential, corresponding to the ground state, where self gravity of the condensate is matched by an effective pressure due to the Uncertainty Principle. The solitons found in the simulations have flat cored density profiles that have been shown to accurately match the known time independent solution  of the Schr\"{o}dinger-Poisson equation \citep{Schive2014,Schive2014b,Schive2016}, for which the soliton mass scales inversely with its radius. Furthermore, a scaling relation between the mass of this soliton and its host virial mass has been uncovered by the $\psi$DM simulations, $m_{soliton} \simeq m_{halo}^{1/3}$, such that a more compact dense soliton should be found in more massive galaxies \citep{Schive2014b} and can be understood from the virial relation \citep{Veltmaat2018}.

On scales larger than the de Broglie scale, the evolution of structure in $\psi$DM simulations is indistinguishable from CDM simulations, starting from the same initial conditions, as desired given the well established agreement between CDM and the statistics of large scale structure and the CMB. Hence, although the light bosons contrast completely with the heavy fermions from super symmetry, they actually provide a very viable non-relativistic explanation for the observed coldness of dark matter.

These distinctive and unique soliton related predictions are very interesting as they can readily falsify the Bose-Einstein interpretation for the DM. Here we test this vulnerable prediction against the best available dynamical date for the ultra-diffuse galaxy DF44, and show how, in this $\psi$DM context, the velocity dispersion profile of DF44 is related to the smaller and somewhat less massive classical dwarfs. 

The most important parameter in the $\psi$DM  context is the boson mass, $m_\psi$, that previously has been estimated to be approximately,  $m_\psi\simeq 10^{-22}$ eV \citep{Schive2014}, by identifying the large DM dominated cores observed in dSph galaxies as solitons. This value has been subsequently supported by independent analyses of other dSph galaxies \citep{Chen2017, Broadhurst2019}. With this value we can normalise the $\psi$DM simulations and predict the absolute values of soliton properties as a function of halo mass for comparison with the observations.

Here, we compare our wave-DM predictions for the newly discovered class of UDG galaxies, in particular for DF44 the currently best-studied example, in the Coma Cluster. It was discovered with the pioneering Dragon Fly multi-beam telescope, built for the purpose of reaching unprecedentedly low surface brightness in ground based surveys  \citep{Abraham2014}. The extended stellar profiles of these UDG galaxies and their low surface brightness seem to challenge models of galaxy formation in standard CDM where large tidal forces and ram pressure stripping, or high rotation dark halos have been proposed by \citet{Liao2019} and \citet{Tremmel2019} and even stars formed in outflows by \citet{DiCintio2017}. These ideas are hotly debated and hard to extend to the discoveries of isolated examples of UDGs some of which show modest ongoing star formation and also extended HI.

We also compare DF44 with wave-DM profiles fitted to other well-studied galaxies of higher and lower mass, for which extended velocity dispersions have been measured. In section \S2 we describe our baseline model consisting of the solitonic core plus a NFW-like outer profile; and in \S3 we compare our predictions to the dataset of the Dragon Fly 44 dwarf galaxy, together with other newly discovered dwarf galaxies. We also show the self-consistency of our baseline model for explaining the dispersion profile of those galaxies. Throughout the paper we assume a standard cosmology \citep{Planck16}.

Gas outflows have been appealed to for flattening the central DM density profile of dwarf galaxies in the context of CDM, with increasingly detailed modelling to help capture the complexities of supernova-driven outflows powered by repeated  bursts of star formation, \citep{NFW1996, Gelato1999,Binney2001,Gnedin2002,Mo2004,Read2005,Mashchenko2006,Mashchenko2008}, and also dynamical friction of inflowing clumped baryons with repeated outflows \citep{El-Zant2001,Weinberg2002,Tonini2006,Romano-Diaz2009,Goerdt2010,Cole2011}. The formation of large cores is now questioned in the latest detailed high resolution simulations by \citet{Bose2019} indicating the gravitational coupling between gas outflows and the central DM profile may not be sufficient to significantly modify the central DM profile, with confirmation by  \citet{Benitez2019} who emphasise that only the high gas density regime is relevant for core formation in the CDM context, requiring frequently repeated outflows with gas continuing to dominate the central potential for Gyrs.


Constraints on the boson mass of $\psi$DM have been claimed using the power spectrum of Ly-$\alpha$ forest clustering, by analogy with Warm-DM, assuming the transmission power spectrum of tenuous, highly ionized Hydrogen turns over at a scale set by the boson mass that has been assumed analogous to the turnover predicted by the free streaming scale of Warm-DM. Relatively high boson masses are inferred this way, $\gtrsim 10^{-21}eV$, from the turn over scale observed in the forest in particular at $z>5$ \citep{Irsic2017,Rogers2021}, which under predict the kpc scale cores of dSph galaxies. These estimates do rely on several simplifying assumptions including universal parameterizations for gas heating evolution and a uniform UV radiation from star forming sources after an instantaneous transition to reionization. It has been cautioned that plausible variation of these assumptions can accommodate Warm-DM with a colder, early IGM \citep{Garzilli2017,Garzilli2021}. We emphasise that forest predictions for Wave-DM are currently missing cosmological simulations  incorporating gas hydrodynamics, and that it is already clear that rich substructure is predicted at least for the DM in this context from the full density modulation caused by self-interference of the wave function on the de Broglie scale, pervading all galaxy halos and filaments \citep{Schive2014}. This inherent substructure may be expected to affect the early gas distribution particularly in the epoch before reionization is completed where dense filaments predate galaxy formation in this context, the epoch of which is set by the boson mass. This early filament dominated era is expected to be similar for Warm-DM for which one detailed filament simulation exists, predicting that gas cools to very high density along DM filaments, with the possibility of early star formation\citep{Gao2007}, which can therefore result in a "cosmic dawn" that is very different from LCDM. Furthermore, for Wave-DM the distribution of the first galaxies will be more biased than for CDM, from the absence of low mass galaxies below the Jeans scale that is set by the boson mass \citep{Schive2014b} and hence reionization is expected not only to be later than for CDM but with larger spatial variance, enhancing the forest and 21cm power spectra. Empirical guidance is expected soon with deep JWST imaging into the era of reionization and also from 21cm mapping of early galaxies and filaments traced by stars and HI respectively.

There may be a gathering case for a significant AGN role in early reionization \citep{Padmanabhan2021} as implied by new high redshift $z\simeq 6$ detections of double peaked Ly$\alpha$ emitters  \citep{Hu2016,Bosman2020,Gronke2020}. This adds to claims of unusually wide "gaps" in the forest at higher redshift $z>5$ \citep{Becker2015} and may be taken to indicate late and/or sparsely distributed sources of ionization \citep{Gangolli2021}. These observations lend support to the proposal that AGN are responsible for the bulk of reionization \citet{Madau2015}, which would imply a different heating history and less uniform reionization, with possibly different conclusions regarding the interpretation of the Lyman-forest power spectrum, especially on small scales.

Another heating source for the forest is gas outflows from galaxies that are observed to be ubiquitous at $z>4$ in high-z surveys \citep{Pettini1999, Frye2002}, that will also act to increase spatial and velocity variance on small scales \citep{Oppenheimer2006,Viel2013,Bertone2006} and may explain the apparent lack of damped Ly$\alpha$ absorption near massive high-z galaxies \citep{Adelberger2005}. General gas enrichment by these outflows may also be supported by the common presence of CIV absorption in the forest \citep{Broadhurst2000}, now detected to the lowest detectable column densities with sizeable, $\simeq 0.1$, volume filling factors \citep{Songaila1996,DOdorico2016}.

If with further observations and simulations is it concluded that outflows and AGN heating do not significantly affect the scale or utility of the turnover in the Ly$\alpha$ forest transmission power-spectrum for constraining boson mass, then it is still possible in the context of Wave-DM, with the dominant light boson considered here of $\simeq 10^{-22}$eV, to appeal to a larger initial field misalignment of the axion potential, which has been demonstrated to provide suitable excess small scale power to match the forest data\citep{Leong2019} in the Wave-DM context. Additionally it is also natural to consider adding a subdominant DM contribution from a heavier axion of $\simeq 10^{-20}$ eV to generate more small scale structure, and indeed this does account well for the newly appreciated DM dominated class of Ultra-Faint galaxies orbiting the Milky Way and perhaps also for the common presence of nuclear star clusters in all classes of galaxy \citep{Luu2020}. This multiple axion model is motivated by the discrete axion mass spectrum generically predicted by String Theory \citep{Arvanitaki2010} which may lead to "nested" solitons\cite{Luu2020} and will also boost small scale clustering of DM and gas to a level that may be predicted by future simulations\citep{Hsu2020}.

\section{Dynamical Model of Ultra-Diffuse Galaxies with a Wave Dark Matter Halo}

The light bosons paradigm was firstly introduced by \citet{Widrow1993}, \citet{Sahni2000} and  \citet{Hu2000}, and subsequently re-considered by \citet{Marsh2014, Schive2014, Bozek2015,Hui2017,Veltmaat2018,Robles2019,Niemeyer2019} in relation to the puzzling properties of dwarf spheroidal galaxies. In the simplest version, without self-interaction, the boson mass is the only free parameter, with a fiducial value of $10^{-22}$ eV. 

The first simulations in this context have revealed a surprisingly rich wave-like structure with a solitonic standing wave core, surrounded by a halo of interference that is fully modulated on the de Broglie scale \citep{Schive2014}. The solitonic core corresponds to the ground-state solution of the coupled Schr\"oedinger-Poisson equations, with a cored density profile well-approximated by \citet{Schive2014, Schive2014b}
\begin{equation}\label{eq:sol_density}
\rho_c(r) \sim \frac{1.9~a^{-1}(m_\psi/10^{-23}~{\rm eV})^{-2}(r_{sol}/{\rm kpc})^{-4}}{[1+9.1\times10^{-2}(r/r_{sol})^2]^8}~M_\odot {\rm pc}^{-3}\,,
\end{equation}
Here $m_\psi$ is the boson mass, and $r_{sol}$ is the solitonic core radius, which simulations show scales as halo mass \citep{Schive2014b} in the following way: 
\begin{equation}\label{eq:rc_sol}
    r_{sol} \propto m_\psi^{-1}M_{halo}^{-1/3}\,.
\end{equation}

The simulations also show the soliton core is surrounded by an extended halo of density fluctuations on the de Broglie scale that arise by self interference of the wave function \citep{Schive2014} and is ``hydrogenic" in form \citep{Hui2017,Vicens2018}. These cellular fluctuations are large, with full density modulation on the de Broglie scale \citep{Schive2014} that modulate the amplitude of the Compton frequency oscillation of the coherent bosonic field, allowing a direct detection via pulsar timing \citep{idm2017,idm2018}.

This extended halo region, when azimuthally averaged, is found to follow the Navarro-Frank-White (NFW) density profile \citep{NFW1996, Woo2009,Schive2014, Schive2014b}
so that the full radial profile may be approximated as:
\begin{equation}\label{eq:dm_density}
\rho_{DM}(r) =
\begin{cases} 
\rho_c(x)  & \text{if \quad}  r< 2r_{sol}, \\\\
\frac{\rho_0}{\frac{r}{r_s}\bigl(1+\frac{r}{r_s}\bigr)^2} & \text{otherwise,}
\end{cases}
\end{equation} 
where $\rho_0$ is chosen such that the inner solitonic profile 
matches the outer NFW-like profile at approximately $\simeq 2r_{sol}$, and $r_s$ is the scale radius.
In this context, we can now predict the corresponding velocity dispersion profile by solving the spherically symmetric Jeans equation:
\begin{equation}\label{eq:sol_Jeans}
\frac{d(\rho_*(r)\sigma_r^2(r))}{dr} = -\rho_*(r)\frac{d\Phi_{DM}(r)}{dr}-2\beta\frac{\rho_*(r)\sigma_r^2(r)}{r},
\end{equation}
where $\rho_*(r)$ is the stellar density distribution defined by the standard Plummer profile for the stellar population:
\begin{equation}
\rho_*(r) = \frac{3M_{*}}{4\pi r_{half}}  \biggl(1+\frac{r^2}{r_{half}^2}\biggr)^{-\frac{5}{2}}\,.
\end{equation}
Here, 
$r_{half}$ is the half-light radius, and
$M_{*}$ is the stellar mass.
$\beta$ is the anisotropy parameter, defined as (see \citet{BT}, Equation (4.61))
\begin{equation}\label{eq:sol_beta}
\beta = 1 - \frac{\sigma_t^2 }{\sigma_r^2}.
\end{equation}

Thus, the gravitational potential is given by:
\begin{equation}\label{eq:potbulge1}
d\Phi_{DM}(r) = G \frac{M_{DM}(r)}{r^2} dr\,,
\end{equation}
with a boundary condition  $\Phi_{DM}(\infty)=0$, and the mass enclosed in a sphere of radius $r$ is computed as follows
\begin{equation}\label{eq:massbulge1}
M_{DM}(r) = 4\pi\int_0^r x^2 \rho_{DM}(x)dx\,.
\end{equation}


Finally, to directly compare our predicted dispersion velocity profile with the observations, we have to project the solution of the Jeans equation along the line of sight as follows:
\begin{equation}\label{eq:sol_projected}
\sigma^2_{los} (R) = \frac{2 }{\Sigma(R)} \int_{R}^{\infty} \biggl(1-\beta \frac{R^2}{r^2}\biggr) \frac{\sigma^2_r(r)\rho_*(r)}{(r^2-R^2)^{1/2}} r dr\,\,
\end{equation}
where 
\begin{equation}
\Sigma(R) =2\int_{R}^{\infty} \rho_*(r)(r^2-R^2)^{-1/2}rdr\,. 
\end{equation}

We now apply the above to the newly measured dispersion profile of the recently discovered galaxy ``Dragon Fly 44" (DF44). This unusual galaxy was recently discovered in a Coma cluster survey using the unique multi-beam optical Dragon Fly imager, designed to search for extended low-surface-brightness emission \citep{Dokkum2016}. This unique telescope has successfully surveyed sizeable areas to unprecedentedly low levels of surface brightness from the ground \citep{Abraham2014}, uncovering a surprising, unknown extreme class of ``Ultra Diffuse Galaxies", that are smooth, diffuse clouds of stars found mainly in massive galaxy clusters, that contrast with the centrally sharp and bright elliptical galaxy members \citep{Dokkum2016}. The galaxy DF44 is one of the largest example of this UDG class, with a half-light radius $\sim 4.6$ kpc, and a stellar mass of $\simeq 3\times10^{8} M_{\odot}$. Its stellar velocity dispersion profile has recently been measured with deep spectroscopy \citep{Dokkum2016,Dokkum2019}.

\section{Minimal Wave-Dark-Matter Radial Profile Comparison}

Here we first compare the measured velocity dispersion and light profile of DF44 with Wave-DM with the minimum number of parameters that are consistent with the findings from our simulations and then subsequently we allow a wider range of soliton and halo mass to allow comparison with previous work for more general conclusions. 

 For our minimal model profiles we solve the spherically symmetric Jeans equation, described above, Eq. \eqref{eq:sol_Jeans}, subject to a total mass of $4\times 10^{10}M_\odot$, which is the dynamical mass estimated by \citet{Dokkum2019} for this galaxy adopting the virial estimate commonly used for dwarf spheroidal galaxies \citep{Walker2009}. We also adopt the commonly used Plummer 'profile for the stellar profile and match it the measured half-light projected radius of 4.6 kpc measured for DF44 \citep{Dokkum2019}. 
 
 For our minimal model adopt the soliton--halo mass scaling relation of Eq. \eqref{eq:rc_sol} discovered in the $\psi$DM simulations \citep{Schive2014b}. This provides the scale of the soliton for a fixed total halo mass. The $\psi$DM simulations have also shown that the NFW form provides a good azimuthal description of the granular $\psi$DM halos outside the soliton core. Hence we may fix the scale length of the NFW profle, $r_s$, to provide the total mass computed with our model for a given choice of concentration we set $r_s = 8$ kpc.  All values are summarized in Table \ref{tabla:1}.

 This comparison shows simply that the general level of velocity dispersion measured for DF44 of $\simeq 30~km/s$ is consistent with the widely favoured boson mass from similar analyses of dwarf spheroidal galaxies $\simeq 10^{-22}$ eV, for which the observed absence of any central rise in the velocity dispersion favours a predominance of tangential over radial dispersion. We can also see that for a larger choice of boson mass of $1.2\times10^{-22}$ eV, a larger value of $\beta$ can be tolerated. This contrasts with the significantly larger mean boson mass of $\simeq 3\times10^{-22}$ eV, highlighted by \citet{Wasserman2019} in a recent analysis of DF44, but is consistent with the lower end of their 95\% range of
 $1.2\times10^{-22}$ eV.

\begin{figure}
	\centering
	\includegraphics[width=0.99\columnwidth]{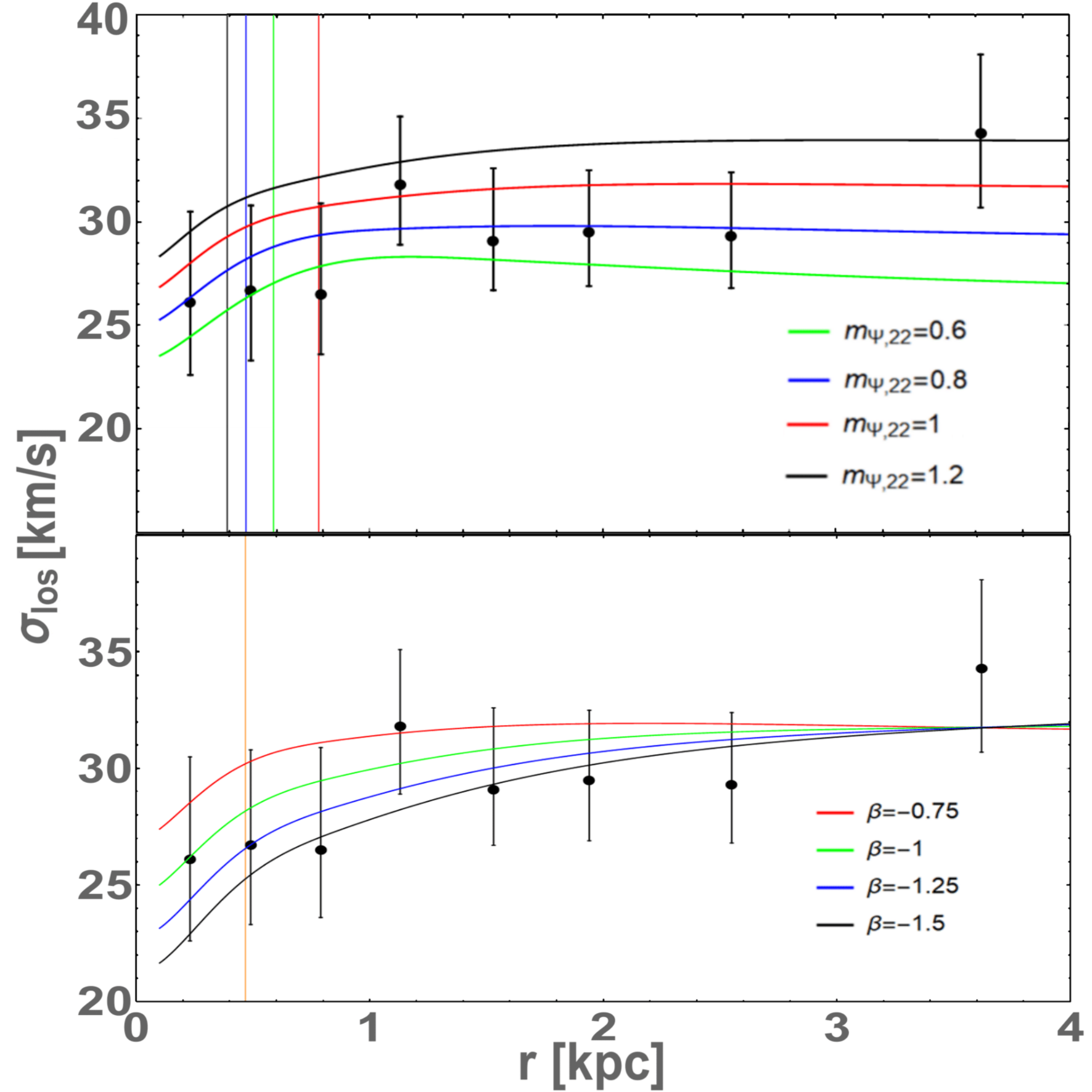}
	\caption{The figure shows the acceptable range of predicted velocity dispersion profiles comparison with DF44 for our minimal model, where the soliton scale is determined by the halo mass and the boson mass. We vary boson mass in the upper panel, setting $\beta=-0.8$ and in the lower panel we set the boson mass to $m_\psi=10^{-22}$ eV and vary $\beta$ according to the legend. This show that a light boson mass in the above range, with modestly negative $\beta$, produces acceptable reduced $\chi^2$. }\label{fig1}
\end{figure}

Acceptable minimal model profiles for DF44 are shown in  Figure \ref{fig1} for several illustrative values for the minimal set of parameters listed in Table \ref{tabla:1}, varying the boson mass and the velocity anisotropy parameter. These profiles have reduced $\chi^2$ values near unity, consistent with the measured dispersion profile of DF44 for its observed large half light radius and virial mass estimate. In the upper panel of Figure \ref{fig1} we vary the boson mass in the range $[0.6, 1.2]\times10^{-22}$ eV, setting $\beta=-0.8$ on the basis of the analysis in \citet{Dokkum2019}. In the bottom panel, we do the opposite by setting $m_\psi=10^{-22}$ eV and varying the anisotropy parameter in the range $[-0.75, -1.5]$. In both cases, the wave-like DM is able acceptably reproduce to the dispersion velocity profile out to $r\sim3$ kpc and even outsider for lower panel profiles. 

\subsection{Comparison of the velocity profile of DF44 with other low and intermediate mass Galaxies}
\label{sec:maths} 

Here we place the velocity profile of DF44 in the wider context of dwarf galaxy profiles for the $\psi$DM framework, spanning the full range of lower masses appropriate for dwarf galaxies. We compare our minimal model with the velocity dispersion profile of representative well studied low mass galaxies whose halo mass and half-light radius span the ranges from $\sim10^9 M_\odot$ to $\sim 5\times10^{10} M_\odot$ and from $\sim 0.8$ kpc to $\sim 5$ kpc, respectively. Since our minimal model must assume a halo mass to compute the corresponding solitonic profile, Eq. \eqref{eq:rc_sol}, we take care to check that the predicted total mass from our model is compatible within the errors with the total mass estimates from observations.

\begin{figure}
	\centering
	\includegraphics[width=0.99\columnwidth]{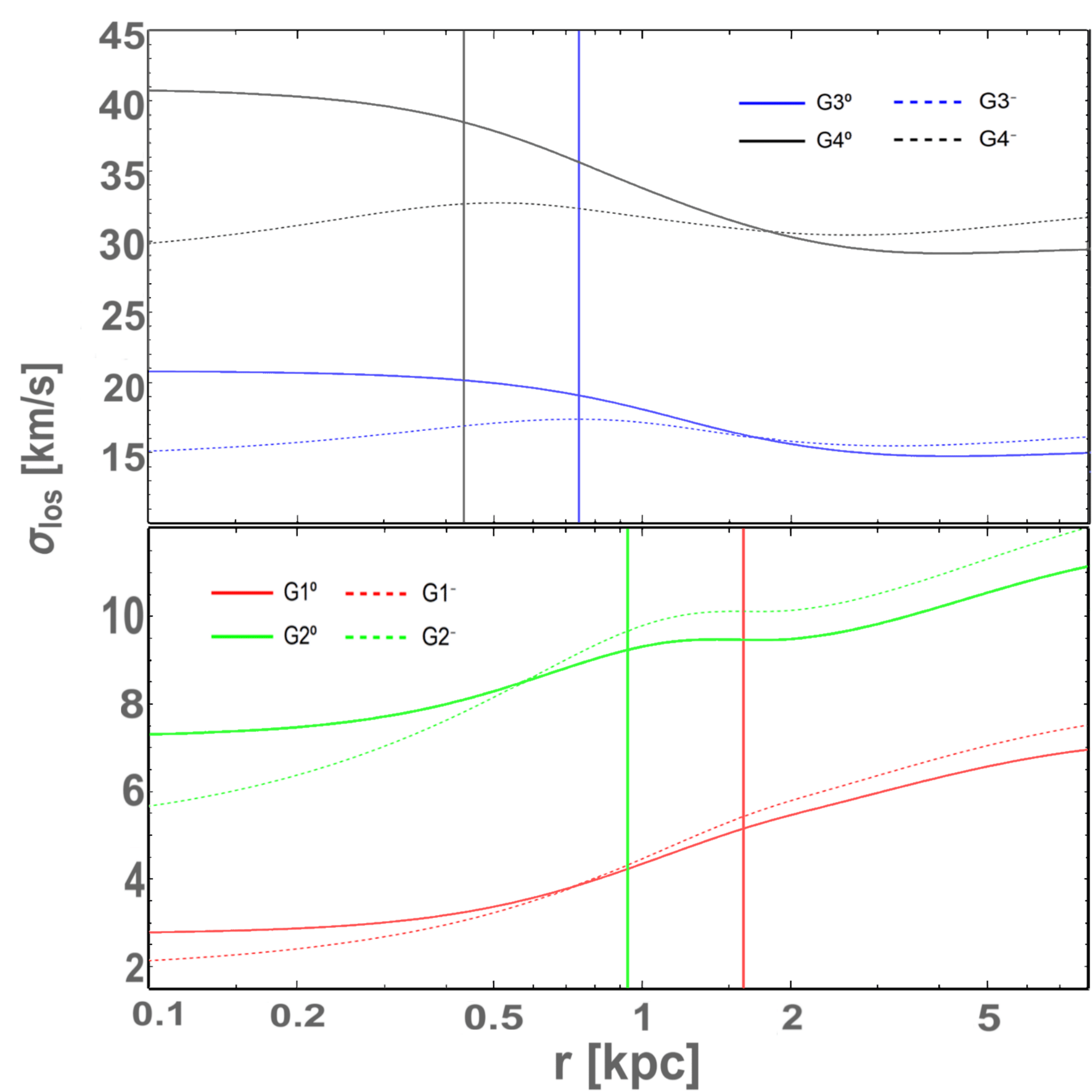}
	\caption{In this figure, we show model solutions for all model galaxies listed in Table \ref{tabla:2} to highlight the transitional feature in the radial profile from the soliton to the NFW-like halo. Solid lines represent solution of the isotropic Jeans equation ($\beta=0$), while dashed lines indicates the predicted dispersion profiles for negative values of $\beta$. For a better visualization, we plot the galaxies with higher mass in the upper panel, while the remaining galaxies are shown in the lower panel.
	}\label{fig2}
\end{figure}

\begin{table}
\centering
\begin{tabular}{|c|c|c|c|c|c|c|c|c|}

\hline
\hline
 Model   & $M_{halo}$  & $r_{sol}$  & $r_{half}$  & $M_{*}$ & $r_s$ & $\beta$ \\
      &($10^{10}M_{\odot}$) &  (kpc) &  (kpc) &  ($10^{7}M_{\odot}$) &  (kpc) & \\
\hline
G1$^0$  & & & & & & 0.0 \\
 & $0.1$ & 1.6 & 0.6 & $0.3$   & 8.5 & \\
G1$^-$  & & & & & & $-0.5$\\
 \hline
G2$^0$  & &  && & &  0.0\\
 & $0.5$ & 0.94 & 0.6 & $0.3$   &  10.5 & \\
G2$^-$  & & & &  && $-0.5$ \\
\hline
G3$^0$  & & & & & & 0.0 \\
 & $1.0$ &0.74 & 2.5 & $3.0$   & 9.0 &  \\
G3$^-$  & & & &  && $-0.5$ \\
\hline
G4$^0$  & & & & & & 0.0 \\
 & $5.0$ & 0.43 & 2.5 & $3.0$   & 7.5 &  \\
G4$^-$  & & & &  && $-0.5$ \\
 \hline
 \hline
\end{tabular}
\caption{Values of the parameters used to construct the velocity dispersion profile of a set of {\em simulated} dwarf galaxies [G1,G2,G3,G4] for two values of the anisotropy parameter. We also vary here the Plummer scale radius for the stars, which allows for a range in "depth" that the stars may occupy in 3D, as listed in Table 1, given by the half light radius. }
\label{tabla:2}
\end{table}
In Figure \ref{fig2} shows illustrative velocity dispersion profiles for a range of $\psi$DM mass profiles highlighting the transition from the soliton core to the outer NFW-like outer profile \citep{Schive2014,Schive2014b,Vicens2018}. The velocity dispersion profile are listed in  Table \ref{tabla:2}, and cover one order of magnitude in the total mass starting from $10^{9} M_{\odot}$ and for different choices for the extent of the stellar profile, and for each model profile compare two representative values of $\beta=[0.0 , -0.5]$ to show the effect of the transition in a family of physical systems representing dwarf galaxies. Solid and dashed lines depict in Fig. 2 for each system, our predicted dispersion profile for $\beta$ set to $0.0$, and $-0.5$, respectively. 

\begin{figure}
	\centering
	\includegraphics[width=0.99\columnwidth]{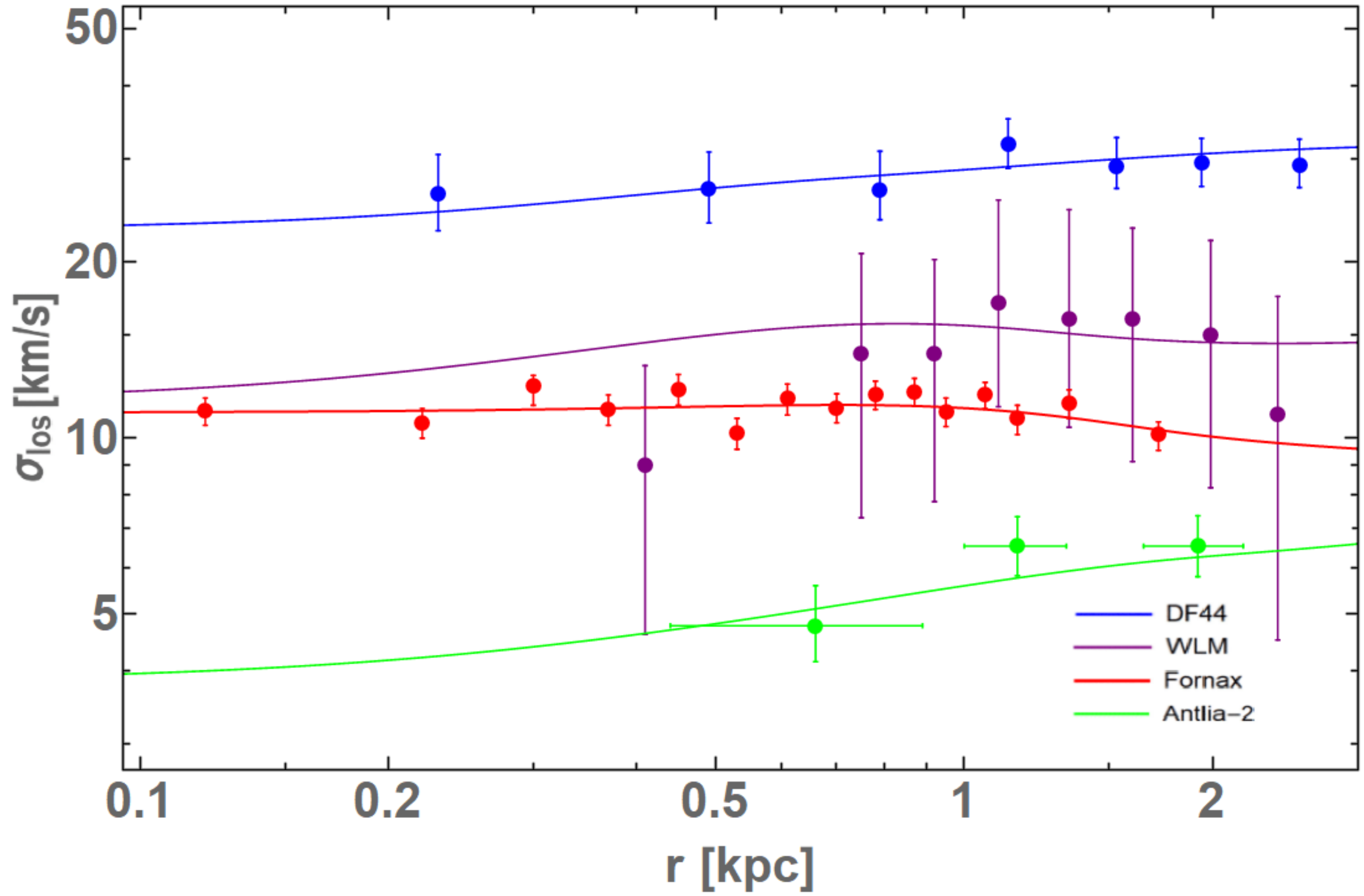}
	\caption{ Comparison of the observed velocity dispersion profiles  with the predictions, for all galaxies listed in Table \ref{tabla:1}.
	}\label{fig3}
\end{figure}

\begin{table*}
\centering
\begin{tabular}{|c|c|c|c|c|c|c|c|c|c|c|}
\hline
\hline
Galaxies  &  $m_\psi$ & $M_{halo, obs}$& $r_{sol}$ & $r_s$   &$r_{half}$  & $M(r<r_{half}),_{obs}$&$M(r<r_{half}),_{th}$ &$\beta$ &$\chi^2_{red}$ & Refs.\\
 & ( $ 10^{-22}$ eV)&($ 10^9 M_{\odot}$)&(kpc) & (kpc)&(kpc)&($ 10^7 M_{\odot}$)&($ 10^7 M_{\odot}$)& & & \\
\hline
DF44 &1 & 40& 0.47&8&$4.6\pm0.2$&$390\pm50$& 300& -1.25& 0.87 &\citet{Dokkum2019} \\ \hline
WLM  & 1& 10& 0.74& 7 &$1.66\pm0.49$&$43\pm3$& 25& -0.75  &0.82 & \citet{Leaman2012} \\ \hline
Fornax & 1& 7& 0.84& 4 &0.85&6.1-8& 6.57& 0.0&1.21&\citet{Mashchenko2015} \\ \hline
 Antlia II& 1  & 1& 1.6 &7&$2.9\pm0.3$&$5.5\pm2.2$& 9.75 & -0.75 &1.34 & \citet{Torrealba2019} \\ \hline 
 \hline 
\end{tabular}
\caption{Observed and estimated magnitudes of the dwarfs galaxies.}
\label{tabla:1}
\end{table*}

Finally, in Figure \ref{fig3}, we apply our model to the dwarf galaxies listed in Table 2, together with the reduced chi-square.We have aimed to illustrate the family of predicted profile shapes to see how of the velocity profile of DF44 may fit in. It can be seen that the relatively flat profile DF44 appears to be continuous with the lower mass galaxies shown here, including Fornax,and Antila II  \citep{Schive2014b,Broadhurst2019} that span the dSph regime and for which we have previously derived boson masses. We have added the more massive, well studied "transition" galaxy, WLM for broader mass coverage. Note, we don't intend to be exhaustive here in this comparison as this would require an understanding of transformative role of tidal effects which are important for most nearby dwarf galaxies orbiting the major local group galaxies, as we have recently shown in \citet{Pozo2020}. Our results show that the minimal model is able to provide a good fit of these dwarf galaxies, where we vary the whole set of parameters, namely ($M_{halo}, r_s, \beta$). For all galaxies, we also compute the total mass within the half-light radius to ensure that it matches the observations. All results are listed in Table \ref{tabla:1}.




\subsection{Pure NFW profile.}

 Here we consider a pure NFW profile for a range of relatively high concentration parameter, $c$, appropriate for relatively low mass galaxies like DF44, ranging over $20<c<40$, guided by the mass-concentration relation and its inherent dispersion derived from CDM simulations. 
 
 Generically of course a centrally rising dispersion profile is predicted for NFW profiles, shown in Figure 4 (and Table 3 labeled as NFW$_1$ and NFW$_2$) as expected given the inherently cuspy density profile of CDM, but is quite unlike the flat profile observed for DF44. A better match DF44 requires a negative anisotropy parameter, with $\beta<-1$ as shown in Figure 4 (and Table 3 labeled as NFW$_3$, NFW$_4$ and NFW$_5$) in order to more than counter the CDM cusp. The NFW$_5$ model has a concentration parameter has been set to the best NFW fit obtained by \citet{Torrealba2019}, while $\beta$ is the same as explored by \citet{Dokkum2019} and this is marginally acceptable with $\beta=-1.25$, in terms of the velocity dispersion profile. However, the scale length predicted for this model $r_s=3.6$kpc is smaller than the measured half light radius of DF44, therefore appearing to be unreasonable, and this is generally the case for the other solutions we have explored here and listed in Table 3, as generically the cooling of gas required to form stars is not expected to produce a stellar distribution that is more concentrated than the dark matter halo as both stars and DM particles behave as test particles. Furthermore, any subsequent "heating" that may have occurred through interactions subsequent to formation may be expected to affect collisionless stars and dark matter equally.

\begin{table}
\centering
\begin{tabular}{|c|c|c|c|c|}
\hline

Profiles &  $M_{halo}$ & $r_s$& c  &$\beta$ \\
&($ 10^9 M_{\odot}$)&(kpc)&&    \\
\hline

$NFW_{1}$ & 10&  2.89 &15&  0   \\ \hline
$NFW_{2}$ & 8& 2.01 & 20&   0 \\ \hline
$NFW_{3}$  & 40& 4.59 & 15 & -2 \\ \hline 
$NFW_{4}$  & 40 & 3.44& 20 &  -3  \\ \hline 
$NFW_{5}$  & 20 & 3.64& 15 &  -1.25  \\ \hline \hline
\end{tabular}
\caption{DragonFly 44 NFW predicted profiles}
\label{tabla:3}
\end{table}

\begin{figure}
	\centering
	\includegraphics[width=0.99\columnwidth,height=6.5cm]{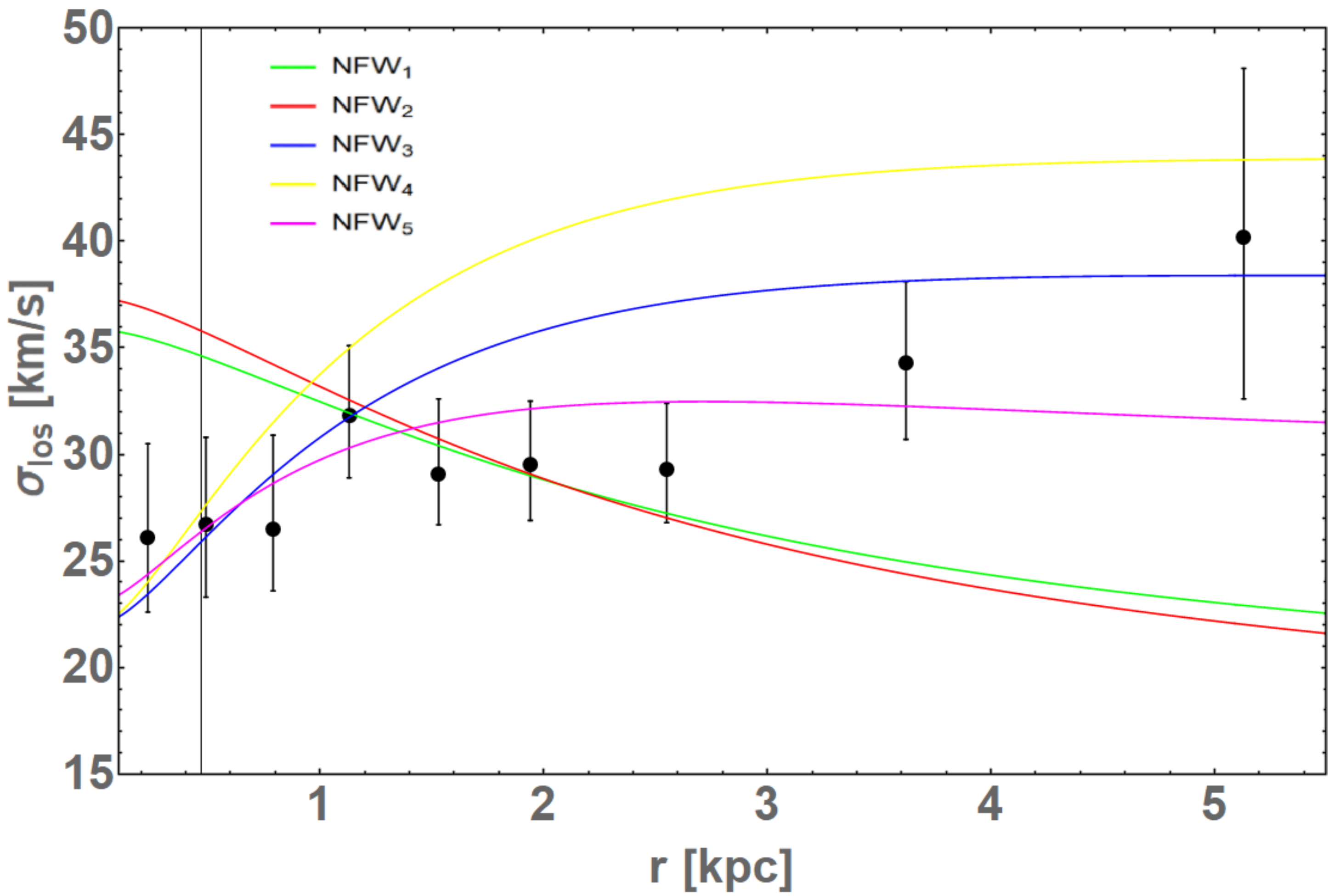}
	\caption{Result of the best pure NFW fits for DF44 listed in table \ref{tabla:3}.The black vertical line is the limit of the calculated soliton for the wave DM profile.
	}\label{fig4}
\end{figure}

\subsection{Generalised Wave-Dark-Matter profile fitting}

Here we explore a fuller range of parameters that does not rely on the "standard" dynamical mass advocated by \citet{Walker2009} for simple stellar systems, since this may not be fully appropriate for the Wave-DM model that we examine here, but has been calibrated in the context of CDM simulations. 

For this purpose we have constructed an MCMC multi-parameter scheme that we compare with the dynamical data and observed scale length for DF44, which we link through the Jeans analysis outlined earlier. In addition to the parameters of the minimal model above, we allow the halo mass to be free and some freedom in "transition radius" between the central soliton and outer NFW halo components given by $\epsilon$ below that is consistent with the spread found in the first simulations \citep{Schive2014}. 


Wide uniform priors are adopted for the above model:
\begin{align}
    -0.5 &< \log_{10}(m_\psi) < 2.5; \\
    1 &< \log_{10}(r_{sol}) < 4; \\
    -1 &< -\log_{10}(1 - \beta) < 1; \\
    2 &< \epsilon < 3.5.
\end{align}
Note, here we have let the matching radius between the soliton and the NFW halo vary over a range of scale, where $r_\text{trans} = \epsilon r_{sol}$, implied by the inherent spread found in the simulations. 

The best fit velocity dispersion curve of the above model is shown in Figure 5, together with the 1-$\sigma$ uncertainty. The preferred value is of the core radius of this model is $\simeq 100$ pc, which is quite small compared to the size of DF44 ($\sim 4.6$ kpc). The preferred halo mass, $1.5^{+3.5}_{-0.7}\times 10^{11}M_\odot$, is several times larger than the dynamical mass adopted earlier of $4\times10^{10}M_\odot$ for our minimal model and the best fitting soliton mass is lower in relation to the halo mass, only 2\% of the halo mass and hence this model is essentially a pure NFW profile with a relatively large tangential anisotropy $\beta\simeq -2$ that counters the inherent NFW cusp, lowering the central velocity dispersion as shown in Figure 5. This halo dominated solution is very similar in terms of halo mass and velocity anisotropy to our best fitting pure NFW profile, NFW$_5$ described in section \S3.2. 

This halo dominated $\psi$DM solution has a somewhat higher favoured boson mass than we derived in section \S 3.1 for our minimal model, but with a sizeable uncertainty: $m_\psi=2.1^{4.9}_{-1.3}\times10^{-22}eV$. This value of the boson mass is compatible with that obtained by \citet{Wasserman2019} for their independent MCMC analysis of DF44 and also agrees with their relatively large negative value of $\beta$. Hence, this higher mass soliton for both the boson mass and the halo mass may be less physically compelling than the minimal model solution we found earlier, for which the reduced $\chi^2$ is certainly acceptable, and for which the soliton is wider so that a relatively less extreme value of $\beta$ is required \citep{Luu2020,Kang2020,Lokas2005,Klimentowski2007}.

\begin{figure}
	\centering
	\includegraphics[width=1\columnwidth,height=7cm]{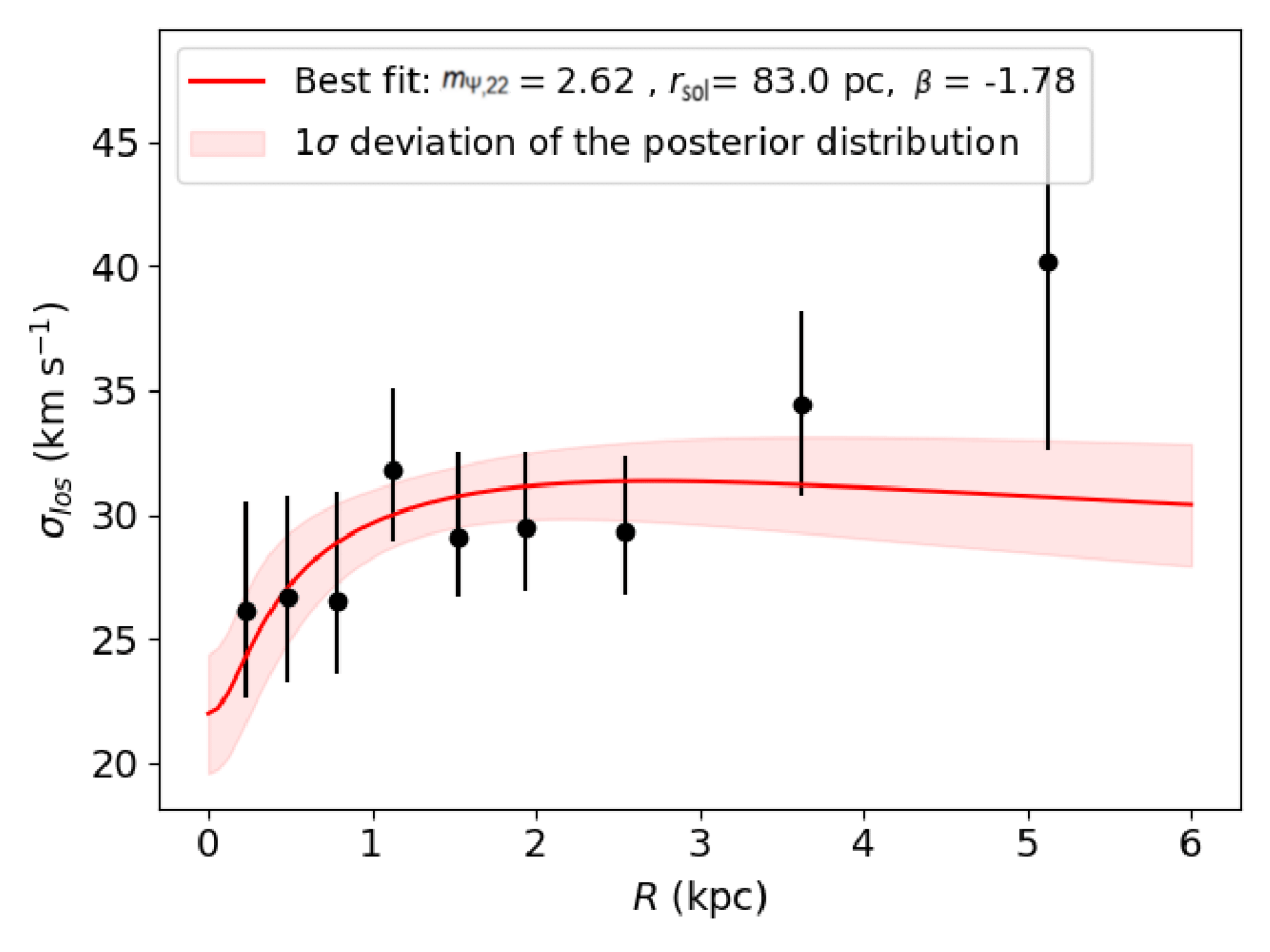}
	\caption{The best fit soliton + NFW model and its one-sigma deviation is calculated by randomly sampling from the distributions of velocity dispersion at each radius.}\label{fig5}
\end{figure}

\begin{figure}
	\centering
	\includegraphics[width=1\columnwidth,height=9cm]{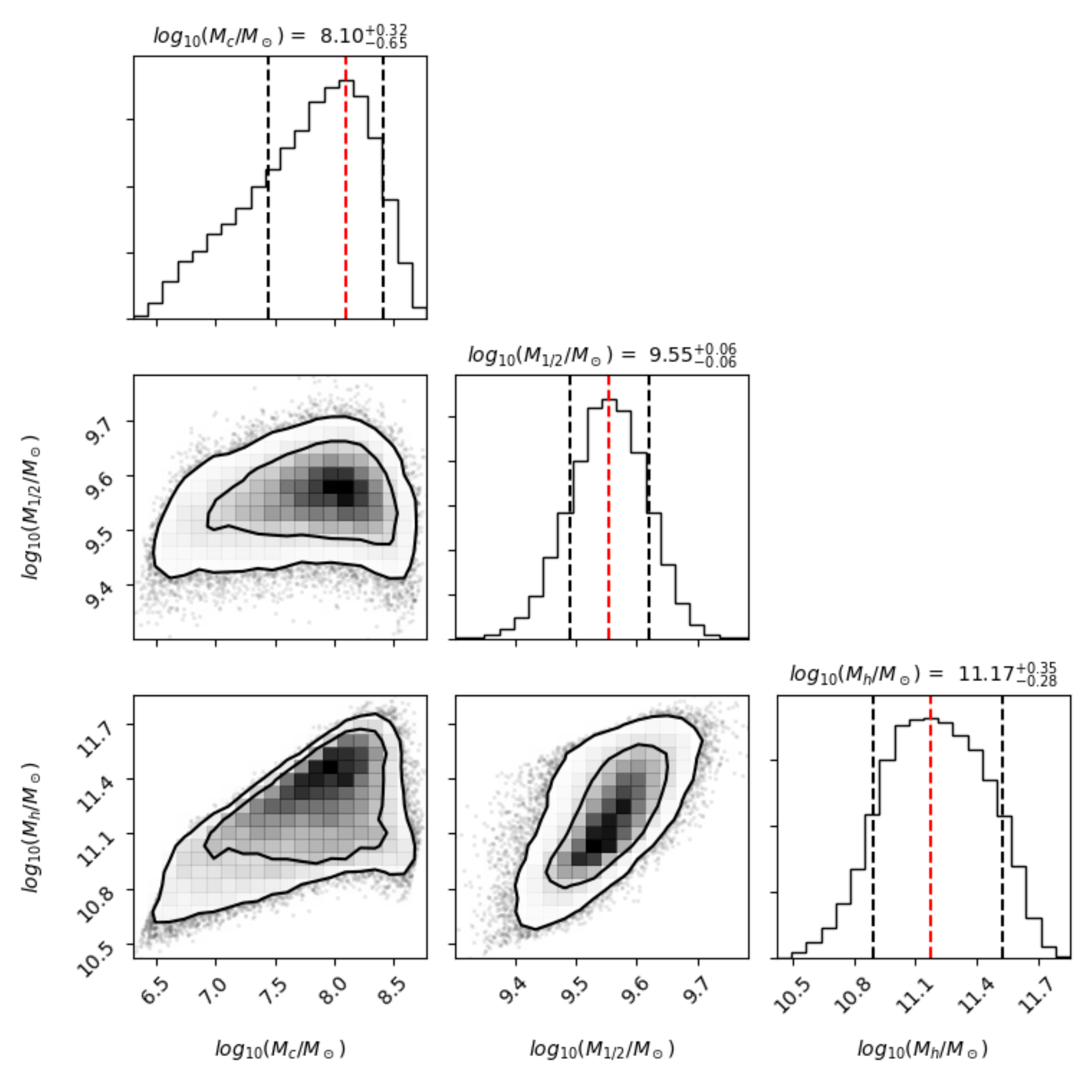}
	\caption{Correlated distributions of inferred parameters: core mass, half-light mass and halo mass from MCMC simulation. Note, the 1D and 2D posterior distributions of four UFDs taken from MCMC chains using emcee \citet{ForemanMackey2013}, plotted using corner package \citet{ForemanMackey2016}}\label{fig6}
\end{figure}

\begin{figure}
	\centering
	\includegraphics[width=1\columnwidth,height=9cm]{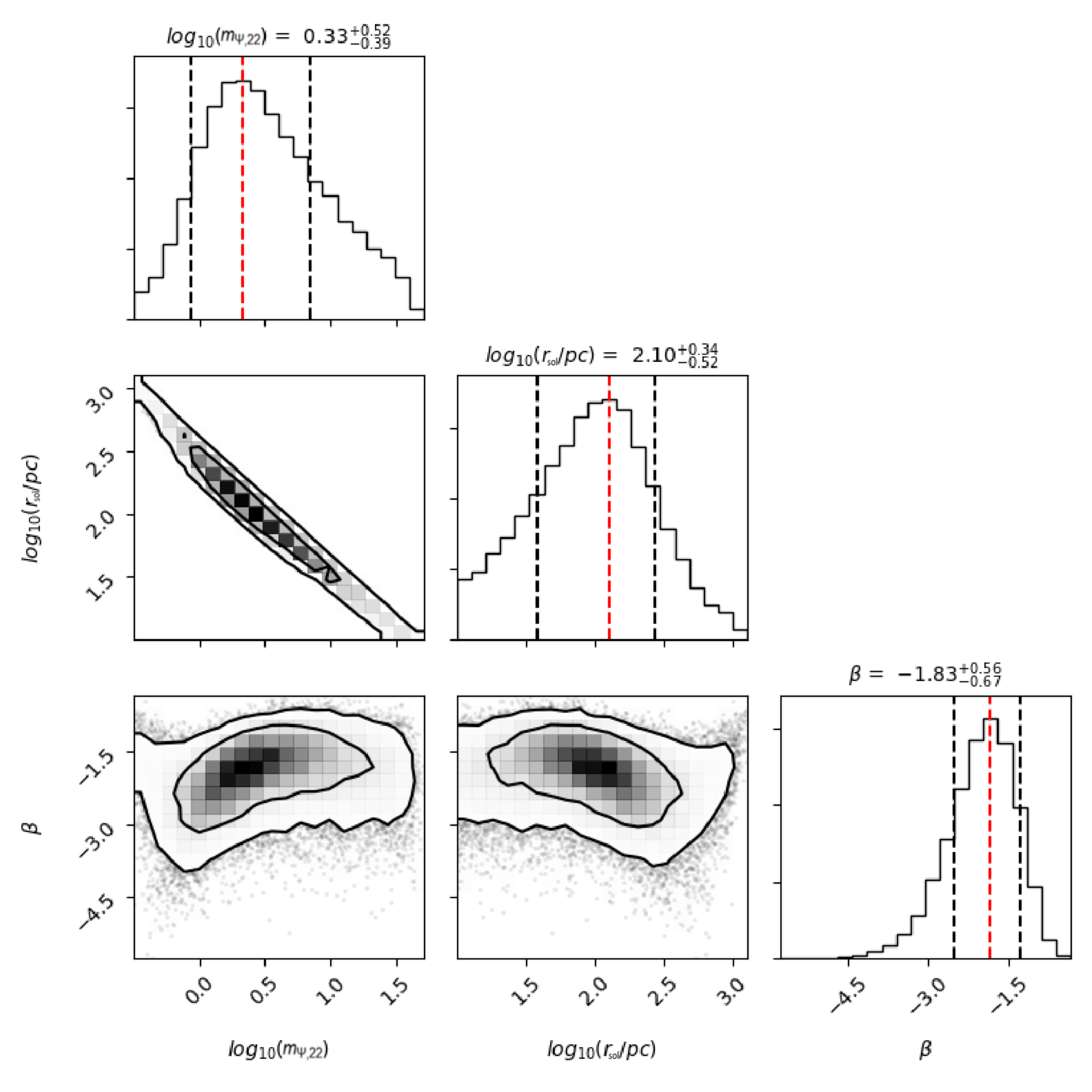}
	\caption{Correlated distributions of free parameters: boson mass, core radius and anisotropy from MCMC simulation. Transition factor distribution is not shown here due to its uniformity. Note, the 1D and 2D posterior distributions of four UFDs taken from MCMC chains using emcee \citet{ForemanMackey2013}, plotted using corner package \citet{ForemanMackey2016}}\label{fig7}
\end{figure}
\section{Discussion and Conclusions}

 Here we have focused on the large low surface brightness galaxy DF44, currently the only example of the newly defined ``ultra diffuse" galaxy class with a measured velocity dispersion profile. A relatively large number of globular clusters appears to be associated with DF44 and initially taken to imply a relatively high mass comparable to the Milky Way, despite the low stellar surface brightness \citep{Dokkum2019}. The new velocity dispersion data, examined here has considerably clarified the mass of DF44, with a velocity dispersion that averages only $ \sigma(r=r_{half}) \simeq 33 \pm 3$ km/s, more indicative of a dwarf galaxy. In terms of morphology UDG galaxies are similar to the dwarf spheroidal galaxies (dSph), however, DF44 differs markedly in terms of the large size of the stellar profile that is several times larger than the $\simeq $kpc stellar scale length of dSph galaxies. 
 
 Here we have been particularly interested in whether this class of galaxy may have implications for the newly appreciated ``Wave-Dark-Matter" interpretation of DM,  because of the  distinctive density profiles found in the first cosmological simulations in this context that predict a rich wave-like structure in the non-linear regime with collapsed halos that should contain solitonic standing wave core of dark bosonic matter at the center surrounded by a wide halo of excited states that are fully modulated in density on the de Broglie scale \citep{Schive2014,Schive2014b,Schive2016}. This characteristic wavelike behaviour has subsequently been verified by other groups independently \citep{Veltmaat2018,Mocz2017} providing a clear distinguishing prediction of rich wavelike structure, unanticipated in the definition of ``Fuzzy Dark Matter"\citep{Hu2000} which we can now see is somewhat of a misnomer, implying an incoherent density profile with a smooth core, quite unlike the rich coherent wavelike structure uncovered in the simulations that is due to pervasive interference on the de Broglie scale with a prominent standing wave core at the center \citep{Schive2014,Schive2014b}. The de Broglie scale sets the scale of soliton radius which is larger for less massive galaxies of lower momentum, as the de Broglie scale is simply set by $\lambda_{dB} = h/(m_\psi v)$, the momentum, given by the level of internal velocity dispersion, and the boson mass, $m_\psi$.

We have shown that this distinctive Wave-DM profile can account for the puzzling combination of the large radius of DF44 and its shallow, low velocity dispersion profile predicted for intermediate mass halos, of approximately $\simeq 4\times10^{10} M_{\odot}$ with a soliton radius of $\simeq 400pc$, this is several times more massive than typical dwarf spheroidal galaxies in the local group neighbourhood for which Wave-DM profiles fitted typically have masses of a few $\times10^{9} M_{odot}$ with a soliton radius of  $\simeq 700pc$  \citep{Schive2014,Chen2017}.  In this intermediate mass range the dispersion profile is relatively flat for DF44, unlike more massive galaxies of high internal momentum, that are predicted to have a narrower, denser central soliton core of approximately 100pc in radius with a correspondingly enhanced central velocity dispersion. This is supported by the rising central velocity dispersion recently measured in the Milky Way \citep{idm2018}. For lower mass galaxies than DF44, wider solitons are predicted, reaching 3 kpc in scale close to the lower limiting ``Jeans mass", below which galaxies do not form in Wave-DM as the dark matter cannot be confined below the de Broglie scale. This limiting soliton size appears to be matched by the newly discovered Antlia II galaxy which is extremely ``ghostly" despite its large size and proximity orbiting the Milky Way and identified by star proper motions with Gaia \citep{Torrealba2019}. For this extreme low density galaxy the velocity dispersion profile is predicted to rise with radius for such a wide soliton to $\simeq 6km/s$ as observed, \citep{Broadhurst2019} and provides a boson mass of $10^{-22}eV$, that is consistent with our analysis here for DF44 and other well studied dSph galaxies, \citep{Broadhurst2019} for a halo of intermediate mass $\simeq 10^{11} M_{\odot}$, consistent with conclusion of \citet{Wasserman2019}. 

Regarding the NFW profile, our Jeans analysis has explored the range of concentrations, $10<c<20$, predicted by $\Lambda$CDM simulations for dwarf galaxies and we have examined the scale length of the Plummer profile used to model the projected star profile for comparison with the measured half light radius of 4.6Kpc for DF44. We firstly adopted the pure NFW halo and isotropic stellar orbits, $\beta=0$, which produced a centrally rising velocity dispersion, that is at odds with the flat profile of DF44, as shown in Figure~3, irrespective of the halo mass. To help flatten the dispersion profile we extend our Jeans analysis to large negative values of $\beta$,  larger than used for $\psi$DM profile and this reverses the predicted velocity dispersion profile, more like the data, as shown in Figure 3, for the profiles with $\beta<0$. To reproduce the large observed stellar half light radius, 4.6kpc requires a high halo mass with a larger mean velocity dispersion as shown as the NFW$_4$ profile in Figure 3. A reasonable compromise can be found by either lowering the concentration allowing a lower halo mass, given by the NFW$_5$ solution listed in Table 3, and shown in Figure 4, for which the DM scale length is  only 30\% smaller than the stellar profile, or by enhancing the velocity anisotropy with the NFW$_3$ solution where the scale length matches the observations with $\beta=-2$. Thus one may conclude that in order to obtain a reasonable fit to DF44 that it is possible to reverse the inherent tendency for a centrally rising profile to match
the observed centrally declining profile by invoking sizeable radial velocity anisotropy.


It is now becoming clear that UDG galaxies like DF44 are surprisingly common in clusters and are also present in the field \citep{Martinez-Delgado2016,Roman2017} that as a class appear to challenge conventional expectations regarding galaxy formation with their large size and low surface brightness. It is also understood that UDG galaxies are more numerous in more massive galaxy clusters \citep{vanderBurg2016,Lee2020} like Coma where they were first recognised by \citet{Dokkum2016} and within the exceptionally massive cluster A2744 where the highest number of UDG galaxies has been detected \citep{Kendall2019}. A interesting connection with HI rich ``HUDS" optically diffuse galaxies has been found by \citet{Leisman2017}, who have identified in the HI selected Alfalfa survey several galaxies with stellar profiles resembling UDG galaxies. Isolated red UDG's have also been discovered ``DGSAT" \citep{Martinez-Delgado2016} and ``S82-DG-1" \citep{Roman2017} but with limited dynamical and distance information. There are also apparently related star forming examples of UDG's  \citep{Prole2019} which resemble the class of local ``transition" dwarf galaxies with stellar velocity dispersion profiles like the dSph's with the additional presence of some rotating HI. Collectively this has been taken to imply that UDG's found in clusters may in general be born in low density environments and either use up their gas over time or subsequently ram pressure stripped of their gas by the hot intra-cluster medium \citep{Liao2019}. 

There is still a dearth of dynamical measurements for UDG galaxies, with DF44 currently the only example reported with a resolved stellar dispersion profile. Improving uncertain distance estimates are also crucial for establishing reliable M/L and radius estimates. The clarity this will bring will then allow the origins of UDG's to clarified in relation to less extreme "normal" galaxies. A unifying picture may be emerging regarding the role of environment, at least for UDGs-like galaxies detected in the HI selected ALFALFA survey, where no preference is found for clusters or underdense regions relative to the general galaxy population \citep{Janowiecki2019} so the surprising properties of UDG galaxies may be intrinsic rather than primarily driven by interaction \citep{Janowiecki2019}. 

We conclude that the NFW is in some significant tension with DF44, but is not firmly excluded provided the relatively large tangential velocity dispersion anisotropy can be deemed reasonable for a NFW halo sufficient to explain the relatively large half light radius scale. Core formation by outflows may help to lower the central dispersion though if this requires repeated outflows of recycled gas for a lasting dark core\citep{Pontzen2012}, including UDG galaxies \citep{Freundlich2020} and a minimum level of star formation \citep{Bose2019,Benitez2019,Dutton2020}this may conflict with the old age and rather low stellar metallicity estimated for DF44\citep{Dokkum2019}. The advantage of $\psi$DM is that it supplies an inherently large core without the need for transformative outflows.

 This consistency we find for DF44 with $\psi DM$ is important as it extends the viability of this form of wave-like DM beyond the dSph class for which clear agreement has been claimed with the wide solitonic cores predicted for this lower mass class of galaxy. The Milky Way provides more evidence of this possibility, as a dense, dark central mass of $10^9M_\odot$ has been uncovered \citep{Portail2017} from the centrally rising dispersion profile of bulge stars that appears in excellent agreement with the expectation of $\psi DM$ for a boson mass of $\simeq 10^{-22}eV$ \citep{idm2020}. 
 
 Further dynamical data and extension to higher mass galaxies should now be explored with high resolution spectroscopy to clarify further where the relatively compact and massive solitons predicted at high mass are supported by the data to establish the basic viability of the coherent light boson hypothesis for the Universal dark matter.

\section*{Acknowledgements}

Alvaro Pozo and Tom Broadhurst are grateful to the Physics department of Hong Kong University and the IAS at HKUST for generous support. George Smoot is grateful to IAS at HKUST for generous support. AP is also thankful for the continued support of the DIPC graduate student program. Also we would like to thank the referee for a thorough reading and enriching comments that have broadened the scope of this work.

\section*{Data Availability}

The data underlying this article will be shared on reasonable
request to the corresponding author.





\bibliographystyle{mnras}

\begin{thebibliography}{99}

\bibitem[\protect\citeauthoryear{Abraham \& van Dokkum}{2014}]{Abraham2014}
Abraham R.,Van Dokkum P., 
 2014, PASP, 126, 55A 
 
 
 
 \bibitem[\protect\citeauthoryear{Adelberger et al.}{2005}]{Adelberger2005}
Adelberger, K. L., Shapley, A. E., Steidel, C. C., Pettini, M., Erb, D. K., Reddy, N. A.
2005, \apj, 629, 636A


   
   
\bibitem[\protect\citeauthoryear{Aprile et al.}{2018}]{Aprile2018}
Aprile, E. \& Xenon Collaboration
2018, \prl, 121k, 1302A 

\bibitem[\protect\citeauthoryear{Archambault et al.}{2017}]{Archambault2017}
Archambault S. \& VERITAS Collaboration
2017, Phys. Rev. D, 95h, 2001A 
   
 
\bibitem[\protect\citeauthoryear{Arvanitaki et al.}{2010}]{Arvanitaki2010}
Arvanitaki, A., Dimopoulos, S., Dubovsky, S., Kaloper, N., March-Russell, J.
2010, Phys. Rev. D, 81l, 3530A
 
  
\bibitem[\protect\citeauthoryear{Becker et al.}{2015}]{Becker2015}
Becker, G. D., Bolton, J. S., Madau, P., Pettini, M., Ryan-Weber, E. V., Venemans, B. P.
2015, \mnras,  447, 3402B 
 
 
 
 


\bibitem[\protect\citeauthoryear{Benitez-Llambay et al.}{2019}]{Benitez2019}
Benítez-Llambay A., Frenk C. S., Ludlow A. D., Navarro J. F.
2019, \mnras,  488, 2387B



\bibitem[\protect\citeauthoryear{Bertone \& White.}{2006}]{Bertone2006}
Bertone, S., White, S. D. M.
2006, \mnras,  367, 247B





\bibitem[\protect\citeauthoryear{Binney,Gerhard \& Silk}{2001}]{Binney2001}
Binney J., Gerhard O., Silk J., 
2001, \mnras,  321, 471B



\bibitem[\protect\citeauthoryear{Binney \& Tremaine}{2008}]{BT}
Binney, J., \& Tremaine, S., 2008, Galactic Dynamics (Princeton, NJ: Princeton Univ. Press) 

\bibitem[\protect\citeauthoryear{Bird et al.}{2016}]{Bird2016}
Bird S.,Cholis I.,Mu\~noz J. B.,Ali-Haïmoud Y.,Kamionkowski M.,Kovetz E. D.,Raccanelli A.,Riess A. G.,
2017, \prl \, 116, t1301B 





\bibitem[\protect\citeauthoryear{Bose et al.}{2019}]{Bose2019}
Bose S.,; Frenk Ca. S., Jenkins A., Fattahi A., Gómez F. A., Grand R. J. J., Marinacci F., Navarro J. F., Oman K. A., Pakmor R., Schaye J., Simpson C. M., Springel V.
2019, \mnras,  486, 4790B



\bibitem[\protect\citeauthoryear{Bosman et al.}{2020}]{Bosman2020}
Bosman, S. E. I., Kakiichi, K., Meyer, R. A., Gronke, M., Laporte, N., Ellis, R. S.
2020, \apj,  896, 49B


\bibitem[\protect\citeauthoryear{Bozek et al.}{2015}]{Bozek2015}
Bozek B.,  Marsh D. J. E.,Silk J.,Wyse R. F.G.,
2015, \mnras,  450, 1, 209B 

\bibitem[\protect\citeauthoryear{Broadhurst et al.}{2019}]{Broadhurst2019}
  Broadhurst T., de Martino I.,Luu  H. N.,  Smoot III G. F., Tye S.-H. H, 
 2020, Phys. Rev. D 101, 083012
  
\bibitem[\protect\citeauthoryear{Broadhurst, Diego \& Smoot}{2018}]{Broadhurst2018}
Broadhurst T., Diego J. M., Smoot G. III,
2018,arXiv:1802.05273B 


\bibitem[\protect\citeauthoryear{Broadhurst \& Scannapieco}{2000}]{Broadhurst2000}
 Broadhurst, T., Scannapieco, E. 
 2000, \apj,  533L, 93B,
 
 
 


  \bibitem[\protect\citeauthoryear{Bullock \& Boylan-Kolchin }{2017}]{Bullock2017}
Bullock J. S., Boylan-Kolchin M.
  2017,Annual Review of Astronomy and Astrophysics, 55, 343B

 
 
 \bibitem[\protect\citeauthoryear{Cappellari et al.}{2005}]{Cappellari2005}
 Cappellari M., Bacon R., Bureau M., Damen M. C.,Davies R. L., de Zeeuw P. T., Emsellem E.,Falcon-Barroso J.,Krajnovic D., Kuntschner H., McDermid R. M.,  Peletier R. F.,Sarzi M.,van den Bosch R. C. E.,van de Ven G.,
2005, \mnras, 366, 1126C 

\bibitem[\protect\citeauthoryear{Chen, Schive \& Chiueh}{2017}]{Chen2017}
Chen S.-R., Schive ., Chiueh T., 
2017, \mnras,  468, 1338C 	
 

\bibitem[\protect\citeauthoryear{Clowe et al.}{2006}]{Clowe2006}
 Clowe D., Bradac M., Gonzalez A.H., Markevitch M., Randall S.W., Jones C.,  Zaritsky D.,2006, \apj, 648, L109-L113 
 
 
 \bibitem[\protect\citeauthoryear{Cole, Dehnen \& Wilkinson}{2011}]{Cole2011}
Cole D. R., Dehnen W., Wilkinson M. I.
2011, \mnras,  416, 1118C 


\bibitem[\protect\citeauthoryear{Cyburt  et al.}{2016}]{Cyburt2016}
 Cyburt R.H.,et al.,2016, Rev. Mod. Phys. 88, 015004 	


\bibitem[\protect\citeauthoryear{Danieli \& van Dokkum}{2018}]{Danieli2018}
Danieli S.,Van Dokkum P.,
2018, \apj, 875, 155D 

\bibitem[\protect\citeauthoryear{de Martino}{2020}]{idm2020}
de Martino I., Broadhurst T., Tye S.-H. H., Chiueh T., Schive  Hsi-Yu, 2020, Physics of the Dark Universe, 28, 100503



\bibitem[\protect\citeauthoryear{de Martino et al.}{2018}]{idm2018}
 de Martino I., Broadhurst T., Tye S.-H. H., Chiueh T., Schive ,
 2018, Galaxies, 6, 10 
\bibitem[\protect\citeauthoryear{de Martino et al.}{2017}]{idm2017}
de Martino I., Broadhurst T.,Tye S.-H. H., Chiueh T., Schive H.-Y, Lazkoz R.,
2017, \prl, 119, 221103 

\bibitem[\protect\citeauthoryear{Di Cintio et al.}{2017}]{DiCintio2017}
Di Cintio A.,Brook C. B., Dutton A. A., Macciò A. V., Obreja A., Dekel A.,
2017, \mnras,466L, 1D


\bibitem[\protect\citeauthoryear{Diego  et al.}{2018}]{Diego2018}
 Diego J.M.,et al.,
 2018, \apj, 857, 25 
 
 
\bibitem[\protect\citeauthoryear{Dutton et al.}{2020}]{Dutton2020}
Dutton, A. A., Buck, T., Macciò, A. V., Dixon, K.i L., Blank, M., Obreja, A.
2020, \mnras,499, 2648D
 
 \bibitem[\protect\citeauthoryear{D'Odorico  et al.}{2016}]{DOdorico2016}
D'Odorico, V., Cristiani, S., Pomante, E., Carswell, R. F., Viel, M., Barai, P., Becker, G. D., Calura, F., Cupani, G., Fontanot, F., Haehnelt, M. G., Kim, T. -S., Miralda-Escudé, J., Rorai, A., Tescari, E., Vanzella, E.,
2016, \mnras,466, 2690D
 
 
 

\bibitem[\protect\citeauthoryear{Ellis}{1984}]{Ellis84}
 Ellis J.,Hagelin J.~S.,Nanopoulos  D.~V., 
Olive K.,Srednicki M., 
1984, Nuclear Physics B., 238, 453E 


\bibitem[\protect\citeauthoryear{El-Zant, Shlosman \& Hoffman}{2001}]{El-Zant2001}
El-Zant Amr., Shlosman I.,Hoffman Y.
 2001, \apj, 560, 636E 
 
 
\bibitem[\protect\citeauthoryear{Foreman-Mackey}{2016}]{ForemanMackey2016}
Foreman-Mackey, D.
2016, JOSS, 1, 24F 
 
  \bibitem[\protect\citeauthoryear{Foreman-Mackey  et al.}{2013}]{ForemanMackey2013}
Foreman-Mackey, D., Hogg, D. W., Lang, D., Goodman, J.
 2013, PASP, 125, 306F 
 
 
  \bibitem[\protect\citeauthoryear{Freundlich et al.}{2020}]{Freundlich2020}
Freundlich, J., Dekel, A., Jiang, F., Ishai, G., Cornuault, N., Lapiner, S., Dutton, A. A., Macciò, A. V.
 2020, \mnras, 491, 4523F 


 \bibitem[\protect\citeauthoryear{Frye, Broadhurst \& Benítez}{2002}]{Frye2002}
Frye B., Broadhurst T., Benítez N.
 2002, \apj, 568, 558F 
 
  
\bibitem[\protect\citeauthoryear{Gangolli et al.}{2021}]{Gangolli2021}
Gangolli, N.,; D'Aloisio, A., Nasir, F., Zheng, Z.
2021, \mnras, 501, 5294G   







 
 \bibitem[\protect\citeauthoryear{Gao \& Theuns}{2007}]{Gao2007}
Gao, L., Theuns, T.
2007, Science, 317, 1527G   
 
 
 
 
 
\bibitem[\protect\citeauthoryear{Garzilli et al.}{2021}]{Garzilli2021}
Garzilli, A., Magalich, A., Ruchayskiy, O., Boyarsky, A.,
2021, \mnras, 502, 2356G   
 
 \bibitem[\protect\citeauthoryear{Garzilli, Boyarsky \& Ruchayskiy}{2017}]{Garzilli2017}
Garzilli, A., Boyarsky, A., Ruchayskiy, O.
2017, Physics Letters B, 773, 258G   
 


\bibitem[\protect\citeauthoryear{Gelato \& Sommer-Larsen}{1999}]{Gelato1999}
Gelato S., Sommer-Larsen J.,
1999, \mnras, 303, 321G  

\bibitem[\protect\citeauthoryear{Gnedin \& Zhao }{2002}]{Gnedin2002}
Gnedin O. Y., Zhao H.
2002, \mnras, 333, 299G  


\bibitem[\protect\citeauthoryear{Goerdt et al.}{2010}]{Goerdt2010}
Goerdt T., Moore B., Read J. I., Stadel J.
2010, \apj, 725, 1707G 




\bibitem[\protect\citeauthoryear{Gronke et al.}{2020}]{Gronke2020}
Gronke, M., Ocvirk, P., Mason, C., Matthee, J., Bosman, S. E. I., Sorce, J. G.; Lewis, J., Ahn, K., Aubert, D., Dawoodbhoy, T., Iliev, I. T., Shapiro, P. R., Yepes, G.
2020, arxiv, 14496G  

  
\bibitem[\protect\citeauthoryear{Gregory et al.}{2019}]{Gregory2019}
 Gregory A. L.,Collins M. L. M.,Read J. I.,Irwin M. J.,Ibata R. A.,  Martin N.F.,McConnachie A. W.,Weisz D. R.,
2019, \mnras, 485, 2010G  

\bibitem[\protect\citeauthoryear{Hu et al.}{2000}]{Hu2000}
Hu W.,Barkana R.,Gruzinov \& A.,
2000, Phys. Rev. Lett., 85, 1158H 




\bibitem[\protect\citeauthoryear{Hu et al.}{2016}]{Hu2016}
Hu, E. M.; Cowie, L. L.; Songaila, A.; Barger, A. J.; Rosenwasser, B.; Wold, I. G. B.
2016, \apj, 825L, 7H 







\bibitem[\protect\citeauthoryear{Hui et al.}{2017}]{Hui2017}
Hui L., Ostriker J. P., Tremaine S., Witten E.,
2017, Phys. Rev. D 95d3541H 



\bibitem[\protect\citeauthoryear{Hsu \& Chiueh.}{2020}]{Hsu2020}
Hsu, Y-H., Chiueh, T.

2020, arXiv, 07602H 





  
\bibitem[\protect\citeauthoryear{Irsic et al.}{2017}]{Irsic2017}
Iršič V., Viel M., Haehnelt M. G., Bolton J. S., Becker G. D.
2017, Phys. Rev. Lett., 119c, 1302I 

   

\bibitem[\protect\citeauthoryear{Islam et al.}{2018}]{Islam2018}
Islam S. ,Rahaman F.,\"Ovg\"un A.,Halilsoy M.,
2018, Canadian Journal of Physics, 97(3): 241-247 

\bibitem[\protect\citeauthoryear{Janowiecki et al.}{2019}]{Janowiecki2019}
Janowiecki S., Jones M. G., Leisman L., Webb A.,
2019, \mnras,490, 566J



\bibitem[\protect\citeauthoryear{Janssens et al.}{2016}]{Janssens2016}
Janssens S, Abraham R, Brodie J.,Forbes D, Romanowsky A. J., Van Dokkum P.,
2017, \apj, 839L, 17J 

\bibitem[\protect\citeauthoryear{Kang et al.}{2020}]{Kang2020}
Kang J-G., Gong Y., Cheng G., Chen X.
2020, Research in Astronomy and Astrophysics, 20, 55K 



\bibitem[\protect\citeauthoryear{Kazantzidis et al.}{2011}]{Kazantzidis2011}
 Kazantzidis S.,Lokas  E. L., Mayer L.,Knebe A.,Klimentowski J. l.,
2011, \apj, 740L, 24K 

\bibitem[\protect\citeauthoryear{Kelly et al.}{2018}]{Kelly2018}
Kelly P.L,.et al.,  
2018, Nature Astronomy, 2, 334-342 

\bibitem[\protect\citeauthoryear{Kendal \& Easther}{2019}]{Kendall2019}
Kendall E., Easther R., 
2020,PASA, 37, 9K



\bibitem[\protect\citeauthoryear{Klimentowski et al.}{2007}]{Klimentowski2007}
Klimentowski J. Łokas E. L., Kazantzidis S., Prada F., Mayer L., Mamon G. A. 
2007, \mnras,378, 353K



\bibitem[\protect\citeauthoryear{Laporte, Agnello \& Navarro }{2018}]{Laporte2018}
 Laporte C. F. P.,Agnello A.,Navarro J. F.,
2018, \mnras,4 84, 245L


\bibitem[\protect\citeauthoryear{Leaman et al.}{2012}]{Leaman2012}
Leaman R., Venn K. A., Brooks A. M., Battaglia G., Cole A. A., Ibata R. A., Irwin M. J.,McConnachie A. W., Mendel J. T., Tolstoy E.,
 2012, \apj, 750, 33L 
 
 
 
 
  \bibitem[\protect\citeauthoryear{Lee et al.}{2020}]{Lee2020}
Lee, J. H., Kang, J., Lee, M. G., Jang, I. S.
2020, \apj, 894, 75L 
 
 
 
 
 
 
 
 
 
 
 \bibitem[\protect\citeauthoryear{Leisman et al.}{2017}]{Leisman2017}
Leisman L., Haynes, M.P., Janowiecki S., Hallenbeck G., J\'ozsa G., Giovanelli R., Adams E. A. K., Bernal Neira D., Cannon J. M., Janesh W. F., Rhode K. L., Salzer J. J.,
2017, \apj, 842, 133L 



\bibitem[\protect\citeauthoryear{Leong et al.}{2019}]{Leong2019}
Leong K-H., Schive H-Y., Zhang U-H., Chiueh T.
2019, \mnras,484, 4273L







 
 
 \bibitem[\protect\citeauthoryear{Liao et al.}{2019}]{Liao2019}
Liao S., Gao L., Frenk C. S., Grand R. J. J., Guo Q., G\'omez F. A., Marinacci F., Pakmor R., Springel S. S.o V.,
2019, \mnras, tmp, 2566L 

\bibitem[\protect\citeauthoryear{Lokas, Mamon \& Prada}{2005}]{Lokas2005}
  Lokas E.L., Mamon G. A., Prada F.
2005, \mnras, 363, 918L  


 
\bibitem[\protect\citeauthoryear{Lokas}{2009}]{Lokas2009}
  Lokas E.L.,
2009, \mnras, 394L, 102L  

\bibitem[\protect\citeauthoryear{Luu \& Broadhurst}{2020}]{Luu2020}
Luu  H. N., Tye S. -H. H., Broadhurst, T.
2020, Physics of the Dark Universe, 3000636L


\bibitem[\protect\citeauthoryear{Madau \& Haardt}{2015}]{Madau2015}
Madau, P., Haardt, F.
2015, \apj,  813L, 8M 

\bibitem[\protect\citeauthoryear{Markevitch et al.}{2004}]{Marikevitch2004}
Markevitch M., Gonzalez A.H., Clowe D., Vikhlinin A., Forman W., Jones C. ,  Murray S.,Tucker W.,
2004, \apj, 606, 819-824 

\bibitem[\protect\citeauthoryear{Marsh \& Silk}{2014}]{Marsh2014}
 Marsh D. J. E.,Silk J., 
2014, \mnras,  437, 2652M 


\bibitem[\protect\citeauthoryear{Martin-Navarro et al.}{2019}]{Martin-Navarro2019}
Martin-Navarro I., Romanowsky A. J., Brodie J. P., Ferre-Mateu A., Alabi A.,  Forbes D. A., Sharina M., Villaume A., Pandya V., Martinez-Delgado D., 
2019, \mnras, 484, 3425M 
 
 \bibitem[\protect\citeauthoryear{Martinez-Delgado et al.}{2016}]{Martinez-Delgado2016}
Mart\'inez-Delgado D, L\"asker R., Sharina M., Toloba E., Fliri J., Beaton R., Valls-Gabaud D., Karachentsev I. D., Chonis T. S., Grebel E. K., Forbes D. A., Romanowsky A. J., Gallego-Laborda J., Teuwen K., G\'omez-Flechoso M. A., Wang J., Guhathakurta P., Kaisin S., Ho N.,
2016, \aj, 151, 96M 
 
\bibitem[\protect\citeauthoryear{Mashchenko}{2015}]{Mashchenko2015}
 Mashchenko S.,
2015, arXiv:1504.08273M 




\bibitem[\protect\citeauthoryear{Mashchenko, Couchman \& Wadsley}{2006}]{Mashchenko2006}
Mashchenko S., Couchman H. M. P., Wadsley J.
2006, Nature  , 442,  7102, 539-542


\bibitem[\protect\citeauthoryear{Mashchenko Wadsley \& Couchman}{2008}]{Mashchenko2008}
Mashchenko S., Wadsley J., Couchman H. M. P.
2008, Science  , 319,  5860, 174

	
\bibitem[\protect\citeauthoryear{Mo \& Mao }{2004}]{Mo2004}
Mo, H. J. Mao, S. 
2004, \mnras, 353, 829M 


\bibitem[\protect\citeauthoryear{Mocz et al.}{2017}]{Mocz2017}
 Mocz P.,et al., 
2017, \mnras, 471, 4559-4570 


\bibitem[\protect\citeauthoryear{Molnar \& Broadhurst}{2018}]{MB2018}
 Molnar S.~M., Broadhurst T., 
2018, \apj,  862, 112M  

\bibitem[\protect\citeauthoryear{Moore}{1994}]{Moore1994}
Moore, B.
1994, Nature, 370, 629M

\bibitem[\protect\citeauthoryear{Navarro, Frenk \& White}{2000}]{NFW1996}
Navarro, J. F., Frenk, C. S.,White, S. D. M.,
1996, \apj,  462, 563N 



\bibitem[\protect\citeauthoryear{Niemeyer}{2019}]{Niemeyer2019}
Niemeyer J. C.,
2020, PrPNP, 11303787N 


\bibitem[\protect\citeauthoryear{Oguri et al.}{2018}]{Oguri2018}
 Oguri M.,Diego J.M.,Kaiser N.,Kelly P.L.,Broadhurst T., 
 2018, \prd, 97, 023518, 




\bibitem[\protect\citeauthoryear{Oppenheimer \& Davé  .}{2006}]{Oppenheimer2006}
Oppenheimer, B. D.; Davé, R.
2006, \mnras, 373, 1265O


\bibitem[\protect\citeauthoryear{Padmanabhan \& Loeb}{2021}]{Padmanabhan2021}
Padmanabhan, H., Loeb, A.
2021, \aa, 646L, 10P 




\bibitem[\protect\citeauthoryear{Pettini et al. }{1999}]{Pettini1999}
Pettini M., 1999, in Walsh J.R., Rosa M.R. 
Chemical Evolution from Zero to High Redshift. Springer-Verlag Berlin, p. 233




\bibitem[\protect\citeauthoryear{Planck Collaboration}{2016}]{Planck16}
Planck Collaboration, 
2016, \aa, {594}, A13 

\bibitem[\protect\citeauthoryear{Prole et al.}{2019}]{Prole2019}
Prole D. J., van der Burg R. F. J., Hilker M., Davies J. I.,
2019, \mnras, 488, 2143P  



\bibitem[\protect\citeauthoryear{Pontzen \& Governato}{2012}]{Pontzen2012}
Pontzen, A., Governato, F.
2012, \mnras, 421, 3464P 





\bibitem[\protect\citeauthoryear{Portail et al.}{2017}]{Portail2017}
 Portail M.,Gerhard O.,; Wegg C., Nes, M.
2017, \mnras, 465, 1621P 


\bibitem[\protect\citeauthoryear{Pozo et al.}{2020}]{Pozo2020}
 
 Pozo A., Broadhurst T., de Martino I., Chiueh T., Smoot G. F., Bonoli S., Angulo R.
2020,   arXiv:2010.10337


\bibitem[\protect\citeauthoryear{Read \& Gilmore }{2005}]{Read2005}
Read, J. I., Gilmore, G.
2005, \mnras, 356, 107R  


\bibitem[\protect\citeauthoryear{Robles et al.}{2019}]{Robles2019}
Robles V. H., Bullock J. S., Boylan-Kolchin M.,
2019, \mnras, 483, 289R  

\bibitem[\protect\citeauthoryear{Rogers \& Peiris}{2021}]{Rogers2021}
Rogers, K. K., Peiris, H. V.
2021, Phys. Rev. Lett., 126g, 1302R   


\bibitem[\protect\citeauthoryear{Rom\'an \& Trujillo}{2017}]{Roman2017}
Rom\'an J.,Trujillo I.,
2017, \mnras, 468, 4039R   



\bibitem[\protect\citeauthoryear{Romano-Diaz et al.}{2009}]{Romano-Diaz2009}
Romano-Díaz E., Shlosman I., Heller C., Hoffman Y.
2009, \apj, 702, 1250R   
    

\bibitem[\protect\citeauthoryear{Ruiz-Lara et al.}{2018}]{Lara2018}
 Ruiz-Lara T.,Beasley M.A.,Falc\'on-Barroso J.,Rom\'an J.,Pinna F.,Brook C.,Di Cintio A.,Mart\'in-Navarro I.,Trujillo I. ,Vazdekis A.,
2018, \mnras, 478, 2034R 


 \bibitem[\protect\citeauthoryear{Sahni \& Wang}{2000}]{Sahni2000}
 Sahni V., Wang L.,
2000, Phys. Rev. D 62j3517S


		
 
 \bibitem[\protect\citeauthoryear{Schive, Chiueh \& Broadhurst}{2014a}]{Schive2014}
 Schive H.-Y.,Chiueh T.,Broadhurst T., 
2014a, Nature Physics , 10,  7, 496-499 

\bibitem[\protect\citeauthoryear{Schive et al.}{2014b}]{Schive2014b}
 Schive H.-Y.,Liao M.-H.,Woo T.-P.,Wong S.-K.,Chiueh T.,Broadhurst T.,   Pauchy Hwang W.-Y.,
2014b, \prl \, 113, 261302 

\bibitem[\protect\citeauthoryear{Schive et al.}{2016}]{Schive2016}
 Schive H.-Y.,Chiueh T.,Broadhurst T.,Huang K.-W., 
2016, \apj, 818, 89  





\bibitem[\protect\citeauthoryear{Songaila \& Cowie}{1996}]{Songaila1996}
Songaila, A., Cowie, L. L.
1996, \aj, 112, 335S  


 
\bibitem[\protect\citeauthoryear{Taibi et al.}{2018}]{Taibi2018}
 Taibi S., Battaglia G., Kacharov N., Rejkuba M., Irwin M., Leaman R., Zoccali M.,  Tolstoy E., Jablonka P.,
2018, \aa, 618A, 122T 


\bibitem[\protect\citeauthoryear {Tonini, Lapi \& Salucci}{2006}]{Tonini2006}
Tonini C., Lapi A., Salucci P.
2006, \apj, 649, 591T  




\bibitem[\protect\citeauthoryear {Torrealba}{2019}]{Torrealba2019}
 Torrealba G., Belokurov V., Koposov S. E., Li T. S., Walker M. G., Sanders J. L., Geringer-Sameth A., Zucker D. B., Kuehn K., Evans N. W., Dehnen W.,
2019, \mnras, 488, 2734T  
    

\bibitem[\protect\citeauthoryear{Tremmel et al.}{2019}]{Tremmel2019}

Tremmel M. Wright A. C., Brooks A. M., Munshi F., Nagai D., Quinn T. R.,
2020, AAS, 23531602T 


\bibitem[\protect\citeauthoryear{van der Burg, Muzzin \& Hoekstra}{2016}]{vanderBurg2016}
van der Burg R.F.J., Muzzin A., Hoekstra H.,
2016, \aa, 590A, 20V 
 


\bibitem[\protect\citeauthoryear{van Dokkum et al.}{2019}]{Dokkum2019}
 van Dokkum P., Wasserman A., Danieli S., Abraham R., Brodie J., Conroy C.,  Forbes D. A., Martin C., Matuszewski M., Romanowsky A. J., Villaume A.,
2019, \apj, 880, 91V 



\bibitem[\protect\citeauthoryear{van Dokkum et al.}{2016}]{Dokkum2016}
 van Dokkum P., Abraham R.,Brodie J.,Conroy C.,Danieli S.,Merritt A.,Mowla L., Romanowsky A., Zhang J.,
2016, \apj, 828L, 6V 




\bibitem[\protect\citeauthoryear{van Dokkum et al.}{2017}]{Dokkum2017}
 van Dokkum P., Abraham R., Romanowsky A. J., Brodie J., Conroy C., Danieli S.,  Lokhorst D., Merritt A., Mowla L., Zhang J.,
2017, \apj, 844L, 11V 
 
 
 \bibitem[\protect\citeauthoryear{Veltmaat et al.}{2018}]{Veltmaat2018}
Veltmaat J., Niemeyer J. C., Schwabe B.,
2018, Phys. Rev. D 98d3509V 

 
 
 \bibitem[\protect\citeauthoryear{Vicens, Salvado \& Miralda-Escudero }{2018}]{Vicens2018}
Vicens J., Salvado J, Miralda-Escudero J.,
2018, arXiv:18021.0513V  







 \bibitem[\protect\citeauthoryear{Viel, Schaye \& Booth}{2013}]{Viel2013}
Viel, M., Schaye, J., Booth, C. M.
2013, \mnras, 429, 1734V  











\bibitem[\protect\citeauthoryear{Walker}{2009}]{Walker2009}
 Walker M. G., Mateo M., Olszewski E. W.. Pe\~narrubia J.. Evans N. W., Gilmore G.,
2019, \apj, 704, 1274W 
  





\bibitem[\protect\citeauthoryear{Wasserman}{2019}]{Wasserman2019}
Wasserman A., van Dokkum P., Romanowsky A. J., Brodie J., Danieli S., Forbes D. A., Abraham R., Martin C., Matuszewski M., Villaume A., Tamanas J., Profumo S.,
2019, \apj, 885, 155W 


\bibitem[\protect\citeauthoryear{Weinberg \& Katz}{2002}]{Weinberg2002}
Weinberg M. D., Katz N.
2002, \apj,  580, 627W 


\bibitem[\protect\citeauthoryear{Widrow \& Kaiser}{1993}]{Widrow1993}
 Widrow L.M.,Kaiser N., 
1993, \apj,  416, 71W 

\bibitem[\protect\citeauthoryear{Woo \& Chiueh}{2009}]{Woo2009}
Woo T-P.,Chiueh T., 
2009, \apj, bf 697, 850W  
















\end{thebibliography}





\bsp	
\label{lastpage}
\end{document}